\documentclass[11pt]{article}
\usepackage{graphicx}
\usepackage{epsfig}
\usepackage[figuresright]{rotating}
\usepackage{amsmath}
\usepackage{amssymb}
\usepackage{amsfonts}

\newcommand{\pibf}{\mbox{\boldmath $\pi$}}
\newcommand{\taubf}{\mbox{\boldmath $\tau$}}

\newcommand{\veceps}{\mbox{\boldmath$\epsilon$}}

\usepackage{cite}
\begin{document}
\begin{center}
{\LARGE{\bf Properties of the $\pi^0$, $\eta$, $\eta'$, $\sigma$,
$f_0(980)$ and $a_0(980)$ mesons and their relevance for the polarizabilities
of the nucleon}}\\[1ex]
Martin Schumacher\\mschuma3@gwdg.de\\
Zweites Physikalisches Institut der Universit\"at G\"ottingen,
Friedrich-Hund-Platz 1\\ D-37077 G\"ottingen, Germany
\end{center}

\begin{abstract}
The signs and values  of the two-photon couplings
$F_{M\gamma\gamma}$ of mesons $(M)$ and their couplings $g_{MNN}$ to the
nucleon as entering into the $t$-channel parts of the difference of the
electromagnetic polarizabilities $(\alpha-\beta)$ and the
backward angle spin polarizabilities $\gamma_\pi$
are determined. 
The excellent agreement achieved with the experimental polarizabilities
of the proton makes it possible to make reliable predictions for the neutron.
The results obtained are
$\alpha_n=13.4\pm 1.0$, $\beta_n=1.8\mp 1.0$ ($10^{-4}$ fm$^3$), and
$\gamma^{(n)}_\pi=57.6\pm 1.8$ ($10^{-4}$ fm$^4$). New empirical information
on the flavor wave functions of the $f_0(980)$ and the $a_0(980)$ 
meson is obtained.
\end{abstract}


\section{Introduction} 

Due to recent research the electromagnetic polarizabilities 
$\alpha$ and $\beta$ and the backward
spin polarizabilities  $\gamma_\pi$ of the proton and the neutron are known
to a level of precision, that details of the $t$-channel contributions
may  be investigated which previously were not of relevance. The $t$-channel
contributions $\gamma^t_\pi$ are given by the pseudoscalar mesons
$\pi^0$, $\eta$, and $\eta'$ whereas the $t$-channel contributions
$(\alpha-\beta)^t$ are given by the scalar mesons $\sigma$, $f_0(980)$,
and $a_0(980)$. The individual contributions depend on the mass $m_M$
of the meson $(M)$, on their two-photon couplings 
$F_{M\gamma\gamma}$ 
and on their couplings  $g_{MNN}$ to the nucleon. In addition to  the 
values of these couplings also  their signs have to be known. 

The $\sigma$-meson enters into the amplitude for Compton scattering as a
$t$-channel exchange. This means that the $\sigma$-meson resonant excited
state is located in the unphysical region of the Compton scattering amplitude
$A_1(s,t)$ at positive $t$. We consider this as an argument in favor of
treating the $\sigma$-meson  contribution to $(\alpha-\beta)^t$ 
in terms of a pole in the complex $t$-plane with properties of the $\sigma$
meson as the particle of the $\sigma$ field, having a definite mass of
$m_\sigma=666.0$ MeV \cite{schumacher06,delbourgo95}. 
However, for the obvious reasons
discussed in \cite{schumacher06,schumacher07} it is equivalent   
to  treat the $\sigma$-meson contribution   to $(\alpha-\beta)^t$     in
terms of a cut in the complex $t$-plane with properties as  obtained for the   
on-shell $\sigma$ meson. This equivalence may be formulated in terms of 
a sum rule,
where $(\alpha-\beta)^t$ connects properties of the $\sigma$-meson
as the particle of the $\sigma$ field with properties of the on-shell
$\sigma$ meson.

The theoretical investigations of the $t$-channel 
pole contributions carried out  in the past 
\cite{schumacher06,schumacher07,lvov99}
remained incomplete because there 
were no firm predictions of the signs. Furthermore, there was only vague
information on the contributions of the  $f_0(980)$
and $a_0(980)$ mesons. The purpose of the present investigation is to provide
the necessary complementary information.

\section{Outline of the problem}

The $t$-channel
contributions entering  into the backward spin polarizability $\gamma_\pi$
and the difference of the electric and magnetic polarizabilities
$(\alpha-\beta)$ may be written in the form 
\begin{eqnarray}
&&\gamma^t_\pi=\frac{1}{2\pi m_N}\left[\frac{g_{\pi
    NN}F_{\pi^0\gamma\gamma}}{m^2_{\pi^0}}\tau_3 
+\frac{g_{\eta
    NN}F_{\eta\gamma\gamma}}{m^2_\eta} 
+\frac{g_{\eta'
    NN}F_{\eta'\gamma\gamma}}{m^2_{\eta'}}\right],
\label{poleeq1}\\
&&(\alpha-\beta)^t=\frac{g_{\sigma NN}F_{\sigma \gamma\gamma}}{2\pi
  m^2_\sigma}
+\frac{g_{f_0 NN}F_{f_0  \gamma\gamma}}{2\pi m^2_{f_0}}
+\frac{g_{a_0 NN}F_{a_0  \gamma\gamma}}{2\pi m^2_{a_0}}\tau_3.
\label{poleeq2}
\end{eqnarray}
In (\ref{poleeq1}) and  (\ref{poleeq2}) $m_N$ is the nucleon mass, $m_{\pi^0}$
{\it etc.} the meson mass, $g_{\pi NN}$ {\it etc.} the meson nucleon coupling
constant, $F_{\pi \gamma\gamma}$  {\it etc.} the meson to two-photon 
coupling and $\tau_3=+1,-1$ for the proton and neutron, respectively.

In addition to the absolute values of the quantities 
entering into (\ref{poleeq1})
and (\ref{poleeq2}) the  signs of the quantities have to be
known. The meson-nucleon coupling constants are known to be positive
quantities whereas the signs of the two-photon couplings  of the mesons have
partly been uncertain. Recently, it has been shown \cite{schumacher06}
that the options
\begin{eqnarray}  
&&F_{\pi^0 \gamma\gamma}=-|M(\pi^0 \to \gamma\gamma)|\\
&&F_{\eta \gamma\gamma}=+|M(\eta \to \gamma\gamma)|\\
&&F_{\eta' \gamma\gamma}=+|M(\eta' \to \gamma\gamma)|\\
&&F_{\sigma \gamma\gamma}=+|M(\sigma \to \gamma\gamma)|
\label{signsofamplitudes}
\end{eqnarray}
lead to good agreement with experiment where the quantities
$M(M\to \gamma\gamma)$
are the meson $M\to \gamma\gamma$ decay  matrix elements. 
In case of the large contributions
corresponding  to the couplings
$F_{\pi^0 \gamma\gamma}$ and $F_{\sigma \gamma\gamma}$, the signs have been
known before from analyses of Compton scattering data,
whereas in case of the small contributions corresponding to the couplings
$F_{\eta \gamma\gamma}$ and $F_{\eta' \gamma\gamma}$ the signs formerly 
have been
adopted as negative \cite{lvov99}. Therefore, it appears desirable to have 
theoretical
arguments which firmly predict the signs in all four cases. These arguments
will be given in the following. In addition also the contributions of the
mesons $f_0(980)$ and $a_0(980)$ will be determined.

The amplitudes for the decay of pseudoscalar mesons $P$ and scalar mesons
$S$ into two photons are given by \cite{donoghue85,deakin94}
\begin{eqnarray}
&&{\cal A}(P\to \gamma(\veceps_1,k_1)\gamma(\veceps_2,k_2))=
M(P\to\gamma\gamma)
\,\epsilon_{\mu\nu\alpha\beta}\,\epsilon^{*\mu}_1\,k^{\nu}_1
\,\epsilon^{*\alpha}_2\,k^\beta_2, \label{amplitude1}\\
&&{\cal A}(S\to
\gamma(\veceps_1,k_1)\gamma(\veceps_2,k_2))=M(S\to\gamma\gamma)
\epsilon_{2\mu}\epsilon_{1\nu}(g^{\mu\nu}\,k_2\cdot k_1 - k^\mu_1\,k^\nu_2)
\label{amplitude2}
\end{eqnarray}
where $\epsilon$ is the polarization vector for linear polarization
and $k$ the 4-momentum of the
photon. Here, the following notation is adopted: 
$g^{00}=-g^{11} =-g^{22}=-g^{33}=1,\quad g^{\mu\nu}=0 \quad 
\mbox{for}\quad \mu\neq\nu, \quad
\epsilon_{\mu\nu\alpha\beta}=1\, \mbox{for even permutations of 0123}, \,
\epsilon_{\mu\nu\alpha\beta}=-1\, \mbox{for odd  permutations of 0123}, 
\,
\epsilon_{\mu\nu\alpha\beta}=0\,\, \mbox{if at least two indices are equal to
each other}$. The kinematical factors on the r.h.s of (\ref{amplitude1})
and (\ref{amplitude2}) for the two-photon decays of the pseudoscalar and
scalar mesons may be evaluated. For this purpose we may write\footnote{
For the phase convention see \cite{yang50}.}
\begin{eqnarray}
&&k_1=(\omega,0,0,k), \label{kinematic1}\\
&&k_2=(\omega,0,0,-k), \label{kinematic2}\\
&&\epsilon_1=(0,1,0,0),\label{kinematic3} \\ 
&&\epsilon_2=(0,1,0,0)\hspace{7mm}\mbox{for scalar mesons},\label{kinematic4}\\
&&\epsilon_2=(0,0,-1,0)\quad\mbox{for pseudoscalar mesons}.\label{kinematic5}
\end{eqnarray}
The different expressions in (\ref{kinematic4}) and (\ref{kinematic5}) are
due to the fact that scalar mesons decay into two photons with parallel
planes of 
linear polarization and pseudoscalar mesons into two photons with 
perpendicular planes of linear
polarization. Inserting  (\ref{kinematic1}) to (\ref{kinematic5}) into
the kinematical factors of  (\ref{amplitude1})  and  (\ref{amplitude2}) we
arrive at 
$\epsilon_{\mu\nu\alpha\beta}\,\epsilon^{*\mu}_1\,k^{\nu}_1
\,\epsilon^{*\alpha}_2\,k^\beta_2=
\epsilon_{2\mu}\epsilon_{1\nu}(g^{\mu\nu}\,k_2\cdot k_1 - k^\mu_1\,k^\nu_2)=-
2\omega^2=-\frac12\,m^2_M$. This means that the kinematical factors in 
 (\ref{amplitude1})  and  (\ref{amplitude2}) are the same except for the fact 
that they distinguish between the two cases of relative linear polarization
of the two photons.

In a quark model the structure of pseudoscalar and scalar may be described as
follows 
\begin{eqnarray}
&& |P\rangle= (a_p |u\bar{u}\rangle + b_p |d\bar{d}\rangle + c_p
|s\bar{s}\rangle)\,(^1S_0),\,\,J^{PC}=0^{-+},\label{wave-1} \\
&& |S\rangle= (a_s |u\bar{u}\rangle + b_s |d\bar{d}\rangle + c_s
|s\bar{s}\rangle)\,(^3P_0),\,\,J^{PC}=0^{++},\label{wave-2} 
\end{eqnarray}
with the constraints $a^2+b^2+c^2=1$, $a^2=b^2$ in both cases. Eqs. 
(\ref{wave-1}) and (\ref{wave-2}) suggest that the flavor parts of the wave
functions may have very similar properties in case of the pseudoscalar 
and scalar mesons, whereas the
angular momentum parts are different. For pseudoscalar mesons we have
$S=0,\,L=0,\,J=0$ and for scalar mesons $S=1,\,L=1,\,J=0$. A further
difference is that the pseudoscalar particles may be considered as
quasistable because of the relatively long lifetimes whereas the scalar mesons
are comparatively shortlived because of the decay of these particles into
meson pairs. This is especially true for the $\sigma$ meson showing up as a
very broad resonance in the $\pi\pi$ channel. As will be shown later, for 
isoscalar mesons we have
$a=b>0$, whereas for isovector mesons we have $-a=b>0$. 
In principle there are further constraints
on the wave functions stemming from $SU(3)$ symmetry. It will turn out
that these constraints lead to very intricate problems with the strange quark
content in case of the scalar mesons. We will introduce here 
a novel way to solve these problems by assuming that these constraints are
strongly violated. The justification for this is that in case of scalar 
mesons the $q\bar{q}$ core state is strongly coupled to two-meson states,
thus leading to a strong distortion of the flavor wave function.
This coupling is known to be also responsible for a major portion of 
the meson mass \cite{vanbeveren86,anisovich06,bugg06,beveren06}. 
The quantum numbers given in (\ref{wave-1}) and 
(\ref{wave-2}) are in line with the fact that pseudoscalar mesons decay 
into two photons with perpendicular planes of linear polarization whereas
scalar mesons decay into two photons with parallel planes of linear
polarization \cite{schumacher05}.

\section{Pseudoscalar mesons}

Using the arguments contained in \cite{close82} ( see p. 27, 28, 54 and 55) 
the flavor wave functions of our present interest may be written in the 
$|{\bf SU(3),SU(2)}\rangle$ form
\begin{eqnarray}
&&|\eta_1\rangle=|{\bf 1,1}\rangle
=\frac{1}{\sqrt{3}}|u\bar{u}+d\bar{d}+s\bar{s}\rangle,
\label{SU3-1}\\
&&|\pi^0\rangle=|{\bf 8,3}\rangle=\frac{1}{\sqrt{2}}|-u\bar{u}+d\bar{d}\rangle,
\label{SU3-2}\\
&&|\eta_8\rangle=|{\bf 8,1}\rangle=\frac{1}{\sqrt{6}}|u\bar{u}+d\bar{d}-2s
\bar{s}\rangle,
\label{SU3-3}
\end{eqnarray}
Using these wave functions  and retaining  $SU(3)$ symmetry
the following relations may be obtained
(see p. 70 of \cite{close82})
\begin{eqnarray}
&&M(\eta_8\to\gamma\gamma)/M(\pi^0\to\gamma\gamma)=-\frac{1}{\sqrt{3}},
\label{Mratio-1}\\
&&M(\eta_1\to\gamma\gamma)/M(\pi^0\to\gamma\gamma)=-2\sqrt{\frac23}. 
\label{Mratio-2}
\end{eqnarray}
The same result for the ratios of matrix elements as given in (\ref{Mratio-1})
and (\ref{Mratio-2}) may be obtained using arguments
contained in \cite{donoghue85,cooper88,scadron04}:
For pseudoscalar mesons  $P$
having the constituent-quark structure
\begin{equation}
|q\bar{q}\rangle={a|u\bar{u}\rangle + b|d\bar{d}\rangle +
 c|s\bar{s}\rangle},\quad a^2+b^2+c^2=1
\label{quarkstructure}
\end{equation}  
the two-photon amplitude may be given in the form 
\begin{equation}
M(P\to\gamma\gamma)= \frac{\alpha_e}{\pi f_P} N_c\, 
\sqrt{2}\,\langle e^2_q \rangle
\label{chargequadrat}
\end{equation}
with
\begin{equation}
\langle e^2_q \rangle={a\, e^2_u + b \,e^2_d
+ c\, (\hat{m}/m_s)\,e^2_s},
\label{equadrat}
\end{equation}
$\alpha_e=1/137.04$, $f_P$  the pseudoscalar decay constant, 
$N_c=3$ the number of colors, $\hat{m}$ the constituent mass of the
light quarks and $m_s$ the constituent mass of the strange quark. 
Numerically we have $m_s/\hat{m}\simeq 1.44$ 
\cite{scadron04,scadron06}.
Defining $C_P=\langle e^2_q\rangle_P/|\langle e^2_q\rangle_{\pi^0}|$
we obtain 
\begin{equation}
C_p={\Big\{} -1,\frac{1}{\sqrt{3}}, 2\sqrt{\frac23} {\Big\}}
\label{Cp}
\end{equation}
for the $\pi^0$, the $\eta_8$ and $\eta_1$ meson, respectively, in the limit
$\hat{m}/m_s\to 1$. This is in agreement with the findings in \cite{donoghue85}
except for the minus sign in front of the first term.
This shows that the ratios given in (\ref{Mratio-1}) and 
(\ref{Mratio-2}) can be obtained from (\ref{chargequadrat}) including the 
signs, and that $M(\pi^0\to\gamma\gamma)$ is negative whereas the two other 
matrix elements are positive. Furthermore, this consideration makes  
transparent that it is the minus sign
in front of the term $u\bar{u}$ in (\ref{SU3-2}), {\rm i.e.}
the flavor component with the larger electric charge,  which leads 
to the minus 
signs in  (\ref{Mratio-1}), (\ref{Mratio-2}) and (\ref{Cp}).

Some additional remarks
concerning the  validity of the minus sign  in front of the term $u\bar{u}$ in
(\ref{SU3-2}) should be made. The signs in (\ref{SU3-1}) to (\ref{SU3-3}) are a
consequence of the use \cite{close82} of the matrix
\begin{equation}
e^{\theta\tau_2/2}=\cos\frac{\theta}{2}+\,i\,\tau_2\sin\frac{\theta}{2}
 \label{positivetau2}
\end{equation}
for carrying out the rotation of a spin-$\frac12$ system through a 
finite angle $\theta$
about the 2-axis in the isospin space, whereas 
\begin{equation}
e^{-\theta\tau_2/2}=\cos\frac{\theta}{2}-\,i\,\tau_2\sin\frac{\theta}{2}
\label{negativetau2}
\end{equation}
as used in other textbooks \cite{halzen84} would lead to  
 $u\bar{u}$ carrying the plus sign  and $d\bar{d}$ carrying  the minus sign
in (\ref{SU3-2}).
In this latter case the three matrix elements $M(\pi^0\to\gamma\gamma)$,
$M(\eta_8^0\to\gamma\gamma)$  and   $M(\eta_1^0\to\gamma\gamma)$ would have the
 same sign. On the other hand we have to realize 
that the 
matrices (\ref{positivetau2}) and (\ref{negativetau2}) are one component
of a 3-vector of matrices for the three axes, so that 
(\ref{negativetau2}) can be obtained from 
(\ref{positivetau2}) through the replacement $\taubf \to -\taubf$ and,
 therefore, also $\tau_3 \to -\tau_3$. As far as the signs in (\ref{poleeq1})
and  (\ref{poleeq2}) are concerned we, apparently, have two equivalent
 options:\\  
{\it Option 1}: We use the sign convention for the pole terms as introduced
in  (\ref{poleeq1}) and  (\ref{poleeq2}) and treat the matrix element
$M(\pi^0\to\gamma\gamma)$ and also $M(a_0\to\gamma\gamma)$ as negative 
quantities
and the other matrix elements  as positive quantities
in accordance with \cite{close82}.\\
{\it Option 2}: We make the replacement $\tau_3\to -\tau_3$
in  (\ref{poleeq1}) and  (\ref{poleeq2}) and treat all the matrix
elements as positive  quantities
in accordance with \cite{halzen84}.\\ 
We will apply  {\it Option 1} throughout this paper.

The physical $\eta$ and $\eta'$ states are defined to be  \cite{bramon99}
\begin{eqnarray}
&&|\eta\rangle=\cos\theta_P|\eta_8\rangle-\sin\theta_P|\eta_1\rangle,
\label{mix1}\\
&&|\eta'\rangle=\sin\theta_P|\eta_8\rangle+\cos\theta_P|\eta_1\rangle,
\label{mix2}
\end{eqnarray}
or equivalently
\begin{eqnarray}
&&|\eta\rangle=\cos\phi_P\frac{1}{\sqrt{2}}|u\bar{u}+d\bar{d}\rangle
-\sin\phi_P|s\bar{s}\rangle,\label{mix3}\\
&&|\eta'\rangle=\sin\phi_P\frac{1}{\sqrt{2}}|u\bar{u}+d\bar{d}\rangle
+\cos\phi_P|s\bar{s}\rangle,\label{mix4}
\end{eqnarray}
with
\begin{eqnarray}
&&\cos\phi_p=\sqrt{\frac13}\cos\theta_P-\sqrt{\frac23}\sin\theta_p,
\label{mix5}\\
&&\sin\phi_p=\sqrt{\frac23}\cos\theta_P+\sqrt{\frac13}\sin\theta_p.
\label{mic6}
\end{eqnarray}
Furthermore, we have
\begin{equation}
\Gamma_{P\gamma\gamma}=\frac{m^3_P}{64\,\pi}|M(P\to \gamma\gamma)|^2.
\label{Gamma-Matrix}
\end{equation}

The results obtained for pseudoscalar mesons are contained in Tables
\ref{flavordependent} and   \ref{flavordependent-2}. With  these Tables 
we wish to find out to what level of precision the physical $\eta$ and 
$\eta'$ mesons can be represented using only one mixing angle $\phi_P$
or equivalently $\theta_p$.
As we can see in Table   \ref{flavordependent-2} the use of only one mixing
angle does not lead to a good  agreement 
between theory and experiment. The existence of two different mixing angles means that the impurities
in $SU(3)$ symmetry cannot be taken into account only by a rotation involving
only the basic $SU(3)$ flavor states, even after the main known
reason for symmetry breaking, i.e.
the $\hat{m}/m_s$ mass ratio in (\ref{equadrat})  
has been taken into account explicitly.

The investigation of the quark structure of pseudoscalar mesons 
and the properties of $\eta- \eta'$ mixing have a long history 
(see e.g. \cite{feldmann00}). The previous analyses cited in  \cite{feldmann00}
led to results ranging from $\theta_p= - 10^\circ$ to $-20^\circ$, i.e. to 
a range of results between the  smaller and the larger value of 
our present analysis.
When trying to solve the problem of the uncertainty of the mixing angle
$\theta_p$ on theoretical grounds the final conclusion was \cite{feldmann00} 
that the $\eta-\eta'$ mixing 
cannot be adequately described by a single mixing angle $\theta_p$,
in agreement with our present result. 
Furthermore, a new mixing scheme is described in \cite{feldmann00}
based on $\chi$PT. In this scheme  the difference between the 
two angles is determined by the difference  
of the pion and the kaon decay constant.  This implies that the connection
between bare octet and singlet states and physical $\eta$ and $\eta'$ states
is not a simple rotation.
\begin{table}[t]
\caption{
The quantity $\langle e^2_q\rangle_P$ is calculated according to 
(\ref{equadrat}) from the flavor wavefunctions. The quantity
$F_{P\gamma\gamma}\equiv M(P\to \gamma\gamma)$ is the corresponding
matrix element for two-photon decay.} 
\begin{center}
\begin{tabular}{c|cccc|}
\hline
&&&&\\
1& meson&$\langle e^2_q\rangle_P$&
$M(P\to\gamma\gamma)$& \\
&&&&\\
\hline
&&&&\\
2&$\pi^0$&
$  -\frac{1}{\sqrt{2}}\frac13$ & 
$-\frac{\alpha_e}{\pi f_\pi}$& \\ 
&&&&\\
3& $\eta$ &
$\frac{5}{9\sqrt{2}}(\cos\,\phi_P-\frac{\sqrt{2}}{5}\frac{\hat{m} }{m_s}
\sin\,\phi_P)$&
$\frac{\alpha_e}{\pi f_\pi}3\sqrt{2}\langle e^2_q\rangle_{\eta}$&\\  
&&&&\\
4&$\eta'$ &
$\frac{5}{9\sqrt{2}}(\sin\,\phi_P+ \frac{\sqrt{2}}{5}\frac{\hat{m} }{m_s}
\cos\,\phi_P)$&
$\frac{\alpha_e}{\pi  f_\pi}3\sqrt{2}\langle e^2_q\rangle_{\eta'} $&\\
&&&&\\
\hline
\end{tabular}
\label{flavordependent}
\end{center}
\end{table}
\begin{table}[b]
\caption{
The meson mass  $m_P$ and the
two-photon decay widths $\Gamma_{P\gamma\gamma}$ are taken from
experiments. The quantity $F^{\rm exp.}_{P\gamma\gamma}$ is
identical  with  $M(P\to \gamma\gamma)$ as given in 
(\ref{Gamma-Matrix}). In line 2 the quantity 
$F^{\rm theor.}_{P\gamma\gamma}$ is calculated form the corresponding
expressions
for     $M(P\to \gamma\gamma)$   in Table \ref{flavordependent}.
In lines 3 and 4 the adjustable parameters are given which lead to agreement
between the theoretical expression and the experimental value for
$F_{P\gamma\gamma}$. }
\begin{center}
\begin{tabular}{c|cllll|}
\hline
&&&&&\\
1& meson&$m_P$&$\Gamma_{P\gamma\gamma}$&$F^{\rm exp.}_{P\gamma\gamma}$ &
$F^{\rm theor.}_{P\gamma\gamma}$ \\
&&&&&\\
&& [MeV]&  [keV]& [$10^{-2}\times$GeV$^{-1}$]& [$10^{-2}\times$GeV$^{-1}$] \\
&&&&&\\
\hline
&&&&&\\
2&$\pi^0$&134.98&$(7.74\pm 0.55)\times10^{-3}$&$-2.52\pm 0.09$& $-2.513\pm
0.007$\\ 
&&&&&\\
3& $\eta$ &547.51& $ 0.510\pm 0.026$ &$2.50\pm 0.06$&$\phi_P=43.1^\circ\pm 
1.0^\circ$\\  
&&&&&$\theta_p=-11.6^\circ\pm 1.0^\circ$\\
&&&&&\\
4&$\eta'$ &957.78& $4.28  \pm 0.19$ & $3.13\pm
0.07$&$\phi_P=36.1^\circ\pm1.4^\circ$\\ 
&&&&&$\theta_p=-18.6^\circ \pm 1.4^\circ$\\
&&&&&\\
\hline
\end{tabular}
\label{flavordependent-2}
\end{center}
\end{table}
\clearpage
\newpage
A possible further 
explanation is given by the admixture of ``extra'' gluonic states  
to the
``ordinary'' $\bar{q}q$ states \cite{ambrosino07}. Such an admixture is 
possible, depending
on the dynamics that we do not understand well \cite{chanowitz88}.

For our further analysis  the results obtained for the $\eta$ and $\eta'$  
mesons
is very important because it teaches us
that constraints from $SU(3)$ symmetry are broken even in the case of
pseudoscalar mesons. This means that these constraints may be
 disregarded to a large extent for scalar mesons where dimeson states
are supposed
to strongly mix into the $q\bar{q}$ flavor wave functions.

\section{Scalar mesons}%%%%%%%%%%%%%%%%%4

In the following we discuss current approaches to low-mass
scalar mesons and a
convenient way how scalar mesons may be treated in the analyses of their
decays to two photons and their coupling to nucleons. As noted before, 
for the present purpose we do not have to take into account the rather 
complicated structure of the on-shell $\sigma$ meson but  may treat
this particle in terms of the dynamical version of the L$\sigma$M.

\subsection{Scalar mesons in the ideal mixing approach}  %%%%%%%%%4.1

In case of ideal mixing the flavor wave-functions of the neutral scalar
mesons are
\begin{eqnarray}
&&|\sigma\rangle= \frac{1}{\sqrt{2}}|u\bar{u}+d\bar{d}\rangle 
\label{scalarflavor-1}    \\
&&|f_0(980)\rangle= |s\bar{s}\rangle \label{scalarflavor-2}\\
&&|a_0(980)\rangle=\frac{1}{\sqrt{2}}
|-u\bar{u}+d\bar{d} \rangle 
\label{scalarflavor-3}
\end{eqnarray}
The structures of these flavor wave functions apparently are in 
agreement with the sign convention of {\it Option 1}, discussed in section
3 and adopted throughout in this paper. Table  \ref{flavordependent-22}
summarizes the analysis of decay matrix elements of  the scalar mesons 
given in \cite{deakin94,beveren02}. We find agreement between experiment and 
prediction for the $\sigma$ and the $f_0$ meson within the errors but a large 
correction factor of $V_q=0.31\pm 0.05$ in case of the $a_0$ meson. This
correction factor has been related to a special property of the $a_0$ meson.
Though the $a_0$ meson has the same flavor wave function as the $\pi^0$
meson the quark-loop calculation may be  different for the scalar meson in
comparison to the pseudoscalar meson. The following relation is derived
for the $a_0(980)$ meson
\begin{eqnarray}
&&M_{\rm
  quark-loop}=2\xi[2+(1-4\xi)I(\xi)]\frac{\alpha_e}{\pi f_\pi} \equiv
V_q\frac{\alpha_e}{\pi f_\pi},\\
&&I(\xi)=\frac{\pi^2}{2}-2\ln^2\left[\frac{1}{\sqrt{4\xi}}
+\sqrt{\frac{1}{4\xi}-1}\right];\quad \xi\leq \frac14,  
\end{eqnarray} 
where $\xi=(m_q/m_{a_0})^2$. This relation leads to a constituent quark
mass $m_q$ as an adjustable parameter.
The parameter $V_q=1$ corresponds to $m_q=360$
MeV whereas $V_q=0.31\pm 0.05$ corresponds to $m_q=231\pm 10$ MeV.
This means that the small decay matrix element of the $a_0$ meson
may be related to a comparatively small constituent quark mass to be inserted
into the loop calculation. 
%%%%%%%%%%%%%%%%%%%%%%%%%%%%%%%%%%% table-3
\begin{table}[h]
\caption{
The quantity $m_S$ is the scalar meson mass taken from experiments, except for
$m_\sigma$  where the prediction of the dynamical version of the L$\sigma$M 
is given.
The two-photon decay widths $\Gamma_{S\gamma\gamma}$ are taken from
experiments in all cases. In lines 2 and 3  the quantity 
$F^{\rm theor.}_{S\gamma\gamma}$ is calculated from the corresponding
expressions for     $M(S\to \gamma\gamma)$.
In lines 4  the value for adjustable parameter $V_q$  is given which 
leads to agreement
between the theoretical expression and the experimental value. }
\begin{center}
\begin{tabular}{c|clllll|}
\hline
&&&&&&\\
1& meson&$m_S$&$\Gamma_{S\gamma\gamma}$&$F^{\rm exp.}_{S\gamma\gamma}$ &
$M(S\to\gamma\gamma)$ &$F^{\rm theor.}_{S\gamma\gamma}$ \\
&&&&&&\\
&& [MeV]&  [keV]& [$10^{-2}\times$GeV$^{-1}$]&& [$10^{-2}\times$GeV$^{-1}$] \\
&&&&&&\\
\hline
2& $\sigma$&666.0& $3.8\pm 1.5$   & $5.0\pm 1.0$&$\frac{\alpha_e}{\pi f_\pi}
\frac53 $&   $4.19$\\
&&&&&&\\
3& $f_0(980)$&980& $0.29^{+0.07}_{-0.09}$  &$0.79\pm 0.11$&
$\frac{\alpha_e}{\pi
  f_\pi}\sqrt{2}\frac13\frac{\hat{m}}{m_s}$
 &$0.82$\\
&&&&&&\\
4& $a_0(980)$&984.7&$0.30\pm 0.10$ &$-0.79\pm 0.13$&$-\frac{\alpha_e}{\pi f_\pi}
V_q(\xi)$&
$V_q=0.31 \pm 0.05$ \\
&&&&&&\\
\hline
\end{tabular}
\label{flavordependent-22}
\end{center}
\end{table}\noindent

\subsection{Scalar mesons with strange quarks in the $a_0$ and non-strange
  quarks in the $f_0$ flavor wave functions in the diquark approach}
  
The weak point of the ansatz of the foregoing subsection
is that in contrast to the pseudoscalar mesons
$\eta$ and $\eta'$ which both contain strange and nonstrange quarks
the scalar mesons $f_0$ and $a_0$ are believed to contain either only
strange quarks or only nonstrange quarks. Therefore, attempts have been made 
to provide a strange-quark component for the $a_0$ meson and a nonstrange
component for the $f_0$ meson. The most prominent example is the introduction
of a cryptoexotic  diquark structure of these two mesons 
\cite{jaffe77}
\begin{eqnarray}
&&|a_0(980)\rangle=\frac{1}{\sqrt{2}}
|sd\bar{s}\bar{d}-su\bar{s}\bar{u}\rangle,\label{diquark-1}\\
&&|f_0(980)\rangle=\frac{1}{\sqrt{2}}
|sd\bar{s}\bar{d}+su\bar{s}\bar{u}\rangle,\label{diquark-2}
\end{eqnarray}
which has a long and heavily debated history. An argument in favor 
of this diquark structure is that the two mesons $a_0$ and $f_0$ are symmetric
with respect to their strange-quark content. But this does not necessarily
mean that the diquark structure is the only one which may provide the
$a_0$ with strange quarks and the $f_0$ meson with non-strange quarks. 
An other possibility is provided by the coupling of these two mesons to
$K\bar{K}$ meson pairs which is discussed in the next subsection.

\subsection{The scalar mesons $f_0(980)$ and $a_0(980)$ within realistic
  meson-exchange models of the $\pi\pi$ and $\pi\eta$ interactions}

The structure of the scalar mesons $f_0(980)$ and $a_0(980)$ has been
investigated by the J\"ulich group \cite{speth95} within realistic
meson-exchange models of the $\pi\pi$ and $\pi\eta$ interactions. The
formalism developed for the $\pi\pi$ system is consistently extended to the
$\pi\eta$ interaction leading to a description of the $a_0(980)$ as a
dynamically generated threshold effect, which is therefore neither a
conventional $q\bar{q}$  state nor a $K\bar{K}$ bound state.

\subsection{The $f_0(980)$ and $a_0(980)$ as $q\bar{q}$ quarkonia coupled to
dimeson states} 

The peak energies of the $f_0(980)$ and $a_0(980)$ resonances are only few MeV
below the $K^+ K^-$ and $K^0 \overline{K^0}$ thresholds 
($2m_{K^\pm}=987.6$ MeV, $2m_{K^0}=995.4$ MeV).  This makes it likely
that the  $f_0(980)$ and $a_0(980)$  mesons are $q\bar{q}$ 
quarkonia states  strongly
coupled to $K^+ K^-$ and $K^0 \overline{K^0}$ dimeson states.
A detailed discussion
of the $K\bar{K}$ fraction in the $f_0(980)$ and the $a_0(980)$ 
is given in \cite{bugg06}.
To give an idea of the order of magnitude discussed there 
we quote that the $K\bar{K}$ fraction
in the $f_0(980)$ is $\sim 70 \%$ and in the $a_0(980)$ $\sim 35 \%$.
In \cite{anisovich06} K-matrix analyses are carried out leading to further 
arguments which favor the opinion that $f_0(980)$ and $a_0(980)$ are dominantly
$q\bar{q}$ states, with a small $(10 - 20  \%)$  admixture of a $K\bar{K}$
loosely bound component.
A treatment of the coupling of the  $q\bar{q}$ quarkonia to dimeson states 
in terms of explicit models is discussed in \cite{beveren06}.

\subsection{Strong isospin breaking  $a_0(980)-f_0(980)$ mixing}

Achasov et al. \cite{achasov79} describe a mechanism through which there 
should be a strong isospin breaking $a_0(980)-f_0(980)$ mixing. 
This phenomenon is shown to be determined by the strong  
couplings 
of the $f_0(980)$ and $a_0(980)$ mesons with the $K^+K^-$ and
$K^0\overline{K^0}$ channels. 
The origin of isospin breaking is the mass difference between the pairs
$K^+K^-$ and $K^0\overline{K^0}$ \cite{achasov79,hanhart07}. Therefore, the 
$a_0(980)-f_0(980)$ mixing amplitude could be shown \cite{achasov79,hanhart07}
to be especially large in the 8 MeV wide interval between the $K^+K^-$
and $K^0\overline{K^0}$ thresholds, but remaining 
sizable outside this interval.

The concept of strong isospin breaking  $a_0(980)-f_0(980)$ mixing
has also been used to point out that these scalar mesons both have
strange and non-strange  $q\bar{q}$ components \cite{wang04}.
This leads to the ansatz
\begin{eqnarray}
&&|f_0(980)\rangle=\cos\phi\,\,
\frac{1}{\sqrt{2}}|u\bar{u}+d\bar{d}\rangle -\sin\phi\,\, 
|s\bar{s}\rangle,
\label{strongmixing-1}\\
&&|a_0(980)\rangle=\sin\phi'\,\,\frac{1}{\sqrt{2}}|-u\bar{u}+d\bar{d}\rangle
+\cos\phi'\,\, |s\bar{s}\rangle,
\label{strongmixing-1}
\end{eqnarray}
which is of special interest for our data analysis described in the following.

\subsection{Ansatz for the flavor wave functions appropriate for the analysis
  of data} %%%%%%%%%%%%%%4.4

In view of the many different models available for the $f_0(980)$ and
$a_0(980)$ mesons we try to avoid the choice of a very specific model for the
data analysis. Instead, we 
propose to make use of the assumption that the wavefunction
may be expanded  in terms of   $q\bar{q}$
components being  strongly  coupled to $K\bar{K}$  pairs.
This coupling  is supposed to be responsible for the masses of the mesons 
\cite{vanbeveren86,anisovich06,bugg06,beveren06} and may as well
have an impact on the relative sizes of the strange-quark content and the
nonstrange-quark content in these two mesons 
\cite{achasov79,hanhart07,wang04}. 
Furthermore, we choose the option to determine
the relative sizes of the $q\bar{q}$ components
empirically.
Details  will be discussed in the next section. We will show
that within this ansatz the two-photon  couplings of the mesons give rather 
direct information on the flavor content of the states. 

\section{Empirical flavor wave-functions of pseudoscalar and scalar mesons}

We start from Eqs. (\ref{wave-1}) and (\ref{wave-2}) and denote pseudoscalar
mesons $(P)$ and scalar mesons $(S)$ by the common  symbol $(M)$. 
 Under this condition and with $\hat{m}/m_s=1/1.44$
we arrive at
\begin{equation}
M(M\to\gamma\gamma)=\frac{\alpha_e}{\pi f_\pi}\frac13 \sqrt{2}
\Big[4a + b + \frac{\hat{m}}{m_s}c\Big]
\label{wave-3}
\end{equation}
where $\frac{\alpha_e}{\pi f_\pi}=0.02513$ GeV$^{-1}$ and $M\to\gamma\gamma$
denotes a two-photon decay of a pseudoscalar or scalar meson. The relation
between the decay matrix element $M(M\to\gamma\gamma)$ and the two-photon decay
width is given by
\begin{equation}
\Gamma_{M\gamma\gamma}=\frac{m^3_M}{64 \pi}|M(M\to \gamma\gamma)|^2.
\label{twophoton-2}
\end{equation}
\begin{table}[h]
\caption{Wave functions, decay matrix elements and decay widths of scalar and
pseudoscalar mesons. The two-photon width $\Gamma_{M\gamma\gamma}$
are taken from \cite{yao06}.}
\begin{center}
\begin{tabular}{ccccccc}
\hline
& $m_M$&$a$&$b$&$c$&$M(M\to\gamma\gamma)$&$\Gamma_{M\gamma\gamma}$\\
& [Mev]& &&& [$10^{-2}$ GeV$^{-1}] $& [keV]\\
\hline
$\eta$& $547.75$& $0.518$ & $0.518$ & $-0.681$ & $+2.50\pm 0.06$ &
$0.510\pm0.026$ \\
$\eta'$& $957.78$& $0.414$& $0.414$& $0.810$& $+3.13\pm 0.05$& $4.29\pm
0.15$ \\ 
$f_0$& $980.0$ & $0.262$ & $0.262$ & $-0.929$& $+0.79\pm 0.11$
& $0.29^{+0.07}_{-0.09} $\\
$a_0$& $984.7$ & $-0.415$ & $0.415$ &$0.810$& $-0.79\pm 0.13$& $0.30\pm
0.10$ \\ 
\hline
\end{tabular}
\end{center}
\label{tableresults}
\end{table} 
The quantity $\Gamma_{M\gamma\gamma}$
in column 7 of Table \ref{tableresults} is the experimental
two-photon decay width, the quantity $M(M\to\gamma\gamma)$ in column 6 the 
decay matrix element where the number is taken from the experimental 
two-photon decay width and the sign from the prediction of the flavor
structure of the meson. The amplitudes $a$, $b$ and $c$ in Table 
\ref{tableresults} have been obtained by adjusting to the experimental
two-photon decay widths given in (\ref{twophoton-2}) using (\ref{wave-3}).

\section{The couplings of the mesons to the nucleon}

\subsection{The $SU(2)\times SU(2)$ linear sigma model}

In the linear sigma model $L\sigma M$, fermions have 
Yukawa couplings with a 
scalar, isoscalar  field $\sigma$ and a pseudoscalar, isovector field 
$\pibf$ \cite{dalfaro73} given by
\begin{equation}
{\cal L}_{\rm int}=
g\bar{\psi}(\sigma'+i\gamma_5\taubf\cdot\pibf)\psi
\label{su2lagrangian}
\end{equation}
where $\sigma'$ is the $\sigma$ field incorporating the effects of chiral
symmetry breaking (see Eqs. (5.50) and (5.51) in \cite{dalfaro73}).
In the dynamical version of the $L\sigma M$ the meson-quark couplings constants
for the $\pi$ and $\sigma$ meson are given by \cite{delbourgo95,schumacher06}
\begin{equation}
g=g_{\pi qq} =  g_{\sigma qq}=2\pi/\sqrt{N_c}=3.63.
\label{mesonquark}
\end{equation}
Then with the Goldberger-Treiman relation for the chiral limit (cl)
\begin{equation}
g f^{\rm cl}_\pi=M
\label{goldberger}
\end{equation}   
we obtain the constituent-quark mass $M=325.8$ MeV in the chiral limit and via
$m_\sigma^2=(2 M)^2+ m^2_\pi$ with $f^{\rm cl}_\pi=89.8$ MeV
and $m_\pi=138.0$ MeV the $\sigma$-meson mass 
\begin{equation}
m_\sigma=666.0\,\,\, \mbox{MeV}.
\label{sigmamass}
\end{equation}
For the $t$-channel part of $(\alpha-\beta)$ due to the $\sigma$ meson
we obtain
\begin{equation}
(\alpha-\beta)^t_{p,n}= \frac{g_{\sigma NN}F_{\sigma\gamma\gamma}}
{2\pi m^2_\sigma}
=\frac{5 \alpha_e \,g_{\pi NN}}{6\pi^2 m^2_\sigma f_\pi}=15.2
\label{alpha-beta}
\end{equation}
in units of $10^{-4}$ fm$^3$, where $\alpha_e=1/137.04$,  
$g_{\sigma NN}\equiv g_{\pi NN}= 13.168\pm 0.057$, $f_\pi=(92.42\pm 0.26)$
MeV.
The result given in (\ref{alpha-beta}) has been obtained through an
application of the L$\sigma$M and its dynamical version. The justification
for using this result in the interpretation of the electromagnetic
polarizabilities  has been given  
in \cite{schumacher06} and \cite{schumacher07} where it has been shown 
that this result leads to an excellent  agreement with experimental data.
In order to obtain $\alpha^t_{p,n}$ and $\beta^t_{p,n}$ separately use may be
made of $(\alpha+\beta)^t_{p,n}=0$.

%%%%%%%%%%%%%%%%%%%%%%%%%%%%%%%%%%%%%%%%%%%%%%%%%%%%%

\subsection{Generalization to $SU(3)\times SU(3)$}

The properties of meson-baryon coupling constants have been derived for
the pseudoscalar meson octet $\{\pi,K,\eta\}$ and the baryon octet 
$\{N,\Sigma,\Lambda,\Xi\}$ using group theory \cite{deswart63} and was
later adapted
to the scalar nonet  $\{\sigma,a_0,f_0,\kappa\}$ \cite{erkol06}.
A compilation may be found in \cite{dumbrajs83}.
As a summary of the investigations in \cite{erkol06}
we may write down the following 
relations for the coupling constants $g_{MNN}$ of a meson $M$ to the nucleon
$N$ in terms of the flavor wave functions:
\begin{eqnarray}
&&g_{MNN}\Big(\frac{1}{\sqrt{2}}(-u\bar{u}+d\bar{d}) \Big)=g_{\pi NN},
\label{quarkg1}\\
&&g_{MNN}\Big(\frac{1}{\sqrt{2}}(u\bar{u}+d\bar{d}) \Big)=g_{\pi
  NN}(4\alpha_M-1), \label{quarkg2}\\ 
&&g_{MNN}(s\bar{s})=0. \label{quarkg3}
\end{eqnarray}
For the mesons $a_0(980)$ (see (\ref{g-1})) and  
$\eta$, $\eta'$, $f_0(980)$ (see (\ref{g-2})) the meson-nucleon
coupling constants, therefore, may be given in the form
\begin{eqnarray}
&&g_{a_0 NN}=\sqrt{2} |a| g_{\pi NN}, \label{g-1}\\
&&g_{MNN}=\sqrt{2} |a| g_{\pi NN}(4\alpha_M-1), \label{g-2}
\end{eqnarray}
where the quantity $|a|$ is absolute value of the amplitude $a$ 
given in Table \ref{tableresults}. 
The quantity
$\alpha_M$ is given by
\begin{equation}
\alpha_M=\frac{F}{F+D}\label{FD}.
\end{equation}
The SU(3) axial vector coupling constants $F$ and $D$ are
determined by neutron and hyperon beta decay. For pseudoscalar mesons
the following numbers are given  (see \cite{thomas00} p. 189)
\begin{equation}
F\simeq 0.51,\quad D\simeq 0.76, \label{FD-1}
\end{equation}
and for the axial vector coupling constant of the nucleon 
\begin{equation}
g_A=F+D\simeq 1.27. \label{FD-2}
\end{equation}
With these  numbers we obtain
\begin{equation}
\alpha_M=\frac{F}{F+D}\simeq\frac25. \label{FD-3}
\end{equation}
For the pseudoscalar states $\eta_8$ and $\eta_1$ we then obtain
\begin{equation}
g_{\eta_8 NN}=\frac{\sqrt{3}}{5}g_{\pi NN},\quad g_{\eta_1
  NN}=\frac{\sqrt{6}}{5} g_{\pi NN}
\label{lvovexample}
\end{equation}
\begin{table}[h]
\caption{Meson nucleon coupling constants}
\begin{center}
\begin{tabular}{llll}
\hline
&$|a|$&$g_{MNN}/g_{\pi NN}$ & $g_{MNN}$\\
\hline
$\eta$&0.518&$(3/5)\sqrt{2}|a|$& $5.79\pm 0.15$\\
$\eta'$&0.414&$(3/5)\sqrt{2}|a|$& $4.63\pm 0.08$\\
$f_0$&0.262&$(6/5)\sqrt{2}|a|$& $5.8\pm 0.8$\\
$a_0$&0.415&$\sqrt{2}|a|$& $7.7\pm 1.2$\\
\hline
\end{tabular}
\end{center}
\label{couplingresults}
\end{table}
in agreement with the result given in \cite{lvov99}. Therefore, it appears
to be well justified to use $\alpha_M=2/5$ for pseudoscalar mesons whereas for
scalar mesons $\alpha_M=0.55$ has been proposed in \cite{erkol06}
and applied here for $f_0$.
The results obtained for the meson-nucleon coupling constants are given in 
Table \ref{couplingresults}. The errors given to the 
quantities $g_{MNN}$ correspond to the experimental errors of
$\Gamma_{M\gamma\gamma}$ in Table   \ref{tableresults}.

\section{Polarizabilities of the neutron derived from the corresponding 
values of the proton} 

In the following we use the well known experimental polarizabilities  of the
proton and the theoretical results obtained for the proton and the neutron
to make more precise predictions for the neutron than obtained directly from
the experiments.

\subsection{Electromagnetic polarizabilities $\alpha$ and $\beta$}

The electromagnetic polarizabilities consist of a resonant part due to the
main resonances of the nucleon, of a nonresonant part which is mainly
due to electric dipole excitation corresponding to  the 
experimental $E_{0+}$ CGLN amplitude of meson photoproduction and of  the
$t$-channel part which is mainly  due to the $\sigma$ meson. The resonant 
parts have formerly been obtained in two different ways \cite{schumacher07}, 
from the directly measured 
parameters of the resonant excited states obtained from the analysis of the
total photoabsorption of the proton \cite{armstrong72}
and from resonance couplings obtained from
analyses of meson photoproduction data of the proton and the neutron
\cite{arndt02,hanstein98,drechsel99}.  
We believe that for the proton the directly
measured parameters of the resonances are more precise than
those  from the analyses of meson photoproduction data. Therefore, we use the
directly measured resonant data of the proton both for the proton and the
neutron. This procedure is unquestionable except for the $F_{15}(1680)$
resonance where the resonance couplings obtained from meson photoproduction
data are much smaller for the neutron as compared to the proton. However, a
recent measurement of the helicity-dependent photoabsorption cross section 
of the
neutron from 815 to 1825 MeV clearly shows that there is no such  difference
in resonant photoabsorption strength for the proton and the 
neutron \cite{dutz05}. 
Therefore, it appears experimentally justified to use the same prediction
for $\alpha$ and $\beta$ for the proton and the neutron also in case of the
$F_{15}(1680)$ resonance. 

The $t$-channel contributions to $\alpha$ and $\beta$ given in lines 2--4
of Table \ref{elmagtable} are by far dominated by the contribution
from the $\sigma$ meson. The contributions from the $f_0$
meson and the $a_0$ meson cancel  in the case of the proton
but interfere constructively with each other and with the contribution of
the $\sigma$-meson in the case of the neutron. This leads to the conclusion 
that parts
of the difference of the electric and magnetic polarizabilities observed 
for the proton and the neutron are due to the $f_0$ and the $a_0$ meson.
The other part of this difference stems from the nonresonant contribution
in line 11 of  Table \ref{elmagtable}. It is well known that the $E_{0+}$
amplitudes for the proton and neutron in the Born approximation and at pion
photoproduction threshold differ by a factor $(1+\frac{m_\pi}{m_N})\simeq
1.15$. This would lead to a ratio 
$\alpha_n(E_{0+})/\alpha_p(E_{0+})\simeq 1.30$ 
in case the  fraction
$(1+\frac{m_\pi}{m_N})\simeq 1.15$ would equally  be valid for the empirical 
$E_{0+}$ amplitudes and would extend unmodified to higher energies. For the
empirical values 
$\alpha_n(E_{0+})/\alpha_p(E_{0+})\simeq 1.28$ was obtained 
\cite{schumacher07} which is in close agreement with the 
value expected for the Born approximation. For the total nonresonant
parts of the electric and magnetic polarizabilities given
in line 11 of Table \ref{elmagtable}
the value 
$\alpha_n({\rm nonres.})/\alpha_p({\rm nonres.})\simeq 1.25$ was adopted which
preserves the normalizations $(\alpha+\beta)_p=13.9\pm 0.3$ and 
$(\alpha+\beta)_n=15.2\pm 0.5$. 
\begin{table}[h]
\caption{Electromagnetic polarizabilities for the proton and the neutron.
Lines 2--4: $t$-channel contributions of the three neutral scalar mesons.
Lines 5--11: $t$-channel, resonant and nonresonant contributions. Line 12:
Sum of lines 5--11. Line 13: experinmental data \cite{schumacher05}. Line 14:
Weighted average of the data in Lines 12 and 13. The data in lines 12--14 are
normalized to $(\alpha+\beta)_p=13.9\pm 0.3$ and 
$(\alpha+\beta)_n=15.2\pm 0.5$.}
\begin{center}
\begin{tabular}{l|l|ll|ll|}
\hline
1& &$\alpha_p$ & $\beta_p$ &  $\alpha_n$ & $\beta_n$ \\
\hline
2&$t-{\rm chan.}\, \sigma$&$+7.6$&$-7.6$&$+7.6$&$-7.6$\\
3&$t-{\rm chan.}\, f_0$&$+(0.3\pm 0.1)$&$-(0.3\pm 0.1)$&$+(0.3\pm 0.1)$&
$-(0.3\pm
0.1)$  \\
4&$t-{\rm chan.}\, a_0$&$-(0.4\pm 0.1)$&$+(0.4\pm 0.1)$&$+(0.4\pm 0.1)$&
$-(0.4\pm
0.1)$  \\
\hline
5&$t-{\rm channel}$&$+7.5$&$-7.5$&$+8.3$&$-8.3$\\
6&$P_{33}(1232)$&$-1.1$& $+8.3$ & $-1.1$& $+8.3$\\
7&$P_{11}(1440)$&$-0.1$&$+0.3$&$-0.1$&$+0.3$\\
8&$D_{13}(1520)$&$+1.2$&$-0.3$&$+1.2$&$-0.3$\\
9&$S_{11}(1535)$&$+0.1$&$-0.0$&$+0.1$&$-0.0$\\
10&$F_{15}(1680)$&$-0.1$&$+0.4$&$-0.1$&$+0.4$\\
11&nonresonant&$+4.5$&$+0.7$&$+5.6$&$+0.9$\\
\hline
12&total&$12.0\pm0.6$&$1.9\mp 0.6$&$13.9\pm 0.6 \pm 1.0$&$1.3\mp 0.6 \mp 1.0$\\
\hline
13&experiment&$12.0\pm0.6$&$1.9\mp 0.6$&$12.5\pm 1.7$&$2.7\mp 1.8$\\
\hline
14&weighted av.&&&$13.4\pm 1.0$&$1.8\mp 1.0$\\
\hline
\end{tabular}
\end{center}
\label{elmagtable}
\end{table}
Two errors are given  for  the neutron 
data in line 12 of  Table \ref{elmagtable}. The first of these errors
was chosen to be the same as the one for the proton, since the proton 
data serve as a measure for the validity of the procedure. The second errors 
are upper limits of possible additional  systematic or model dependent errors.

We consider the results given in line 12 of Table  \ref{elmagtable} 
as a new set of data for the electromagnetic
polarizabilities of the neutron which  supplements to the existing 
data of  line 13. The final result, therefore, is  the weighted average of 
the neutron data
in lines 12 and 13: 
\begin{center}
\framebox{
$\alpha_n=13.4\pm 1.0$, \quad $\beta_n=1.8 \mp 1.0$.
}
\end{center}
These values are also given in line 14 of Table \ref{elmagtable} and in the
abstract.

\subsection{The backward spin polarizability $\gamma_\pi$}

For the $t$-channel parts of the backward spin polarizabilities of the
nucleons rather firm information is available after the  relative
signs of the three contributions have been determined.  The numbers
obtained for the $t$-channel spin-polarizabilities in lines 2 -- 4 of Table
\ref{spinpoltable} are obtained from the two-photon couplings 
given in
column 5 of Table \ref{flavordependent-2} and the meson--nucleon couplings
given in column 4 of Table \ref{couplingresults}. 
The $s$-channel parts in line 6
have formerly been precisely determined by L'vov \cite{lvov99} and are used
here without modification. Excellent agreement between prediction and 
experiment is obtained for the proton. This gives us confidence that a similar
precision for the agreement between theory and experiment  is also 
given for the neutron. 
\begin{table}[h]
\caption{The backward spin polarizabilities  $\gamma_\pi$ for the proton 
and the neutron. Lines 2--4: $t$-channel contributions from the $\pi^0$,
$\eta$ and $\eta'$ meson. Lines 5,6: Total predicted $t$- and $s$-channel
contributions. Line 7: Sum of lines 5 and 6. Line 9: Weighted average
of lines 7 and 8. }
\begin{center}
\begin{tabular}{llll}
\hline
1&  &$\gamma^{(p)}_\pi$&$\gamma^{(n)}_\pi$\\
\hline
2&$t$-chan. $\pi^0$& $-46.7$&+$46.7$\\
3&$t$-chan. $\eta$&$+1.2$&$+1.2$\\
4&$t$-chan. $\eta'$&$+0.4$&$+0.4 $\\
\hline
5&$t$-channel&$-45.1$&$+48.3$\\
6&$s$-channel&$+(7.1\pm 1.8)$&$+(9.1\pm 1.8)$\\
\hline
7&$\gamma_\pi$ theor.&$-(38.0\pm 1.8)$&$+(57.4\pm 1.8 \pm 0.9)$\\
8&$\gamma_\pi$ exp.&$-(38.7\pm 1.8)$&$+(58.6\pm 4.0)$\\
\hline
9&weighted av.&& $+(57.6\pm 1.8)$\\
\hline
\end{tabular}
\end{center}
\label{spinpoltable}
\end{table}
Nevertheless we take into account a possible systematic or model error
in case of the neutron, given by the second error in the neutron data of line
7. This additional error was chosen as an upper limit of possible estimates.
We consider this result of line 7 as a new value for the neutron
which supplements to  the  existing one of line 8. The weighted average 
of the two 
results given in lines 7 and 8, therefore, is the new final result for the
backward spin polarizability of the neutron. This weighted average is
\begin{center}
\framebox{$\gamma^{(n)}_\pi=+(57.6\pm 1.8)$}
\end{center}
which also is given in line 9 of Table \ref{spinpoltable} and in the abstract.

\section{Summary and Discussion}

In the foregoing we have determined the signs and values of the
two-photon couplings $F_{M\gamma\gamma}$ of the 
pseudoscalar mesons $\pi^0$, $\eta$ and $\eta'$ and the scalar mesons
$\sigma$, $f_0(980)$ and $a_0(980)$ and their couplings to the nucleon
$g_{MNN}$  as entering into  the backward angle spin polarizabilities 
$\gamma_\pi$ and the $t$-channel parts of the electromagnetic
polarizabilities $(\alpha-\beta)$.
The  quantities $F_{M\gamma\gamma}$ and  $g_{MNN}$  have been found to be
positive numbers except for the two-photon couplings $F_{M\gamma\gamma}$
of the $\pi^0$ and $a_0$ mesons which are negative  
within the sign convention of the flavor
wave functions adopted here. Because of the high level of precision
obtained for the proton  rather reliable predictions can be made for the 
neutron, thus improving on the precisions of the polarizabilities of the
neutron, well  within the limits given by the experimental polarizabilities.
 The conclusion we have to draw from these findings is that there are no
non-understood contributions to the polarizabilities. For the structure of the
nucleon the following conclusions are obtained. The $s$-channel parts can be
calculated from the resonant and nonresonant cross sections of the nucleon,
where the nonresonant part may be related to  the meson cloud. 
In the case of the
meson-cloud contributions  70\%  are due to nonresonant  electric-dipole  
excitation. The other 30\% consist of magnetic-dipole excitation and combined
processes consisting simultaneously of nonresonant and resonant excitation
processes, resulting in  two-pion emission in photoabsorption experiments.
The $t$-channel parts
can be understood in terms of pseudoscalar and scalar mesons coupled to the
constituent quarks. The quantity $(\alpha-\beta)^t$ is given by the
scalar mesons which couple to two photons with parallel planes of
linear polarization. The quantity $\gamma^t_\pi$ is given by pseudoscalar
mesons which couple to two photons with perpendicular planes of
linear polarization.


\clearpage
\newpage
\begin{thebibliography}{99}

\bibitem{schumacher06}
M. Schumacher, Eur. Phys. J. A \textbf{30}, 413 (2006);
ERRATUM: M. Schumacher, Eur. Phys. J. A \textbf{32}, 121 (2007)
[hep-ph/0609040].

\bibitem{delbourgo95}
R. Delbourgo, M. Scadron, Mod. Phys. Lett. A \textbf{10}, 251 (1995)
[hep-ph/9910242]; Int. J. Mod. Phys. A \textbf{13}, 657 (1998) 
[hep-ph/9807504]; M. Nagy, M.D. Scadron, G.E. Hite, Acta Physica Slovaca
\textbf{54}, 427 (2004);
M.D. Scadron, M. Nagy, [hep-ph/0507168].  

\bibitem{schumacher07}

M. Schumacher, Eur. Phys. J. A \textbf{31}, 327 (2007) [hep-ph/0704.0200].

\bibitem{lvov99}
A.I. L'vov, A.M. Nathan, Phys. Rev. C \textbf{59},  1064 (1999). 


\bibitem{donoghue85}
J.F. Donoghue, B.R. Holstein, Y.-C.R. Lin,
Phys. Rev. Lett. \textbf{55}, 2766 (1985).

\bibitem{deakin94}
A.S. Deakin, V. Elias, D.G.C. McKeon, M.D. Scadron,
Mod. Phys. Lett. A \textbf{9}, 2381 (1994).

\bibitem{yang50}
C.N. Yang, Phys. Rev. \textbf{77}, 242 (1950).

\bibitem{vanbeveren86}
E. van Beveren, T.A. Rijken, K. Metzger, C. Dullemond, G. Rupp, J.E. Ribeiro,
Z. Phys. C \textbf{30}, 615 (1986); N.N. Achasov, G.N. Shestakov, Phys. Rev.
D \textbf{49}, 5779 (1994); N.A. T\"ornqvist, 
Z. Phys. C \textbf{68}, 647 (1995)
[hep-ph/9504372]; N.A. T\"ornqvist, M. Roos, Phys. Rev. Lett. \textbf{76},
1575 (1996) [hep-ph/9511120]; E. van Beveren, G. Rupp, Eur. Phys. J. C
\textbf{10}, 469 (1999) [hep-ph/9806246]; N.A. T\"ornqvist, Eur. Phys. J C
\textbf{11}, 359 (1999);
E. van Beveren, G. Rupp,
Eur. Phys. J. C \textbf{22}, 493 (2001) [hep-ex/0106077]; A. Deandrea 
{\it et al.}, Phys. Lett. B \textbf{502}, 79 (2001) [hep-ph/0012120];
Yu. S. Surovtsev {\it et al.}, Eur.Phys. J.A. \textbf{15}, 409
(2002); M. Boglione, M.R. Pennington, Phys. Rev. D \textbf{65}, 114010 (2002)
[hep-ph/0203149].

\bibitem{anisovich06}
V.V. Anisovich, Int. J. Mod. Phys. A \textbf{21}, 3615 (2006)
[hep-ph/0510409].

\bibitem{bugg06}
 D.V. Bugg, Eur. Phys. J. C \textbf{47}, 57 (2006) 
[hep-ph/0603089].

\bibitem{beveren06}
E. van Beveren {\it et al.}, Phys. Lett. B \textbf{641}, 265 (2006)
[hep-ph/0606022]. 





\bibitem{schumacher05}

M. Schumacher, Prog. Part. Nucl. Phys. \textbf{55}, 567 (2005) 
[hep-ph/0501167].


\bibitem{close82}

F.E. Close, ``An Introduction to Quarks and Partons'' Academic Press 1979,
Fifth Printing 1982.



\bibitem{cooper88}

S. Cooper, Annu. Rev. Nucl. Part. Sci. \textbf{38}, 705 (1988).


\bibitem{scadron04}

M. Scadron et al., Phys. Rev. D \textbf{69}, 014010 (2004)
[hep-ph/0309109].

\bibitem{scadron06}
M. D. Scadron, R. Delbourgo, R. Rupp, J. Phys. G \textbf{32}, 735 (2006)
[hep-ph/0603196].

\bibitem{halzen84}
F. Halzen, A.D. Martin, Quarks \& Leptons, John Wiley \& Sons (1984).

\bibitem{bramon99}
A. Bramon, R. Escribano, M.D. Scadron,   Eur. Phys. J. C  
\textbf{7}, 271 (1999). 


\bibitem{feldmann00}
T. Feldmann, Int. Journ. Mod. Phys. A \textbf{15}, 159 (2000)
[hep-ph/9907491].

\bibitem{ambrosino07}
F. Ambrosino {\it et al.}, Phys. Lett. B \textbf{648}, 267 (2007)
[hep-ex/0612029]; G. Li, Q. Zhao, C.-H. Chang, (2007) [hep-ph/0701020];
C.E. Thomas, (2007) [hep-ph/0705.1500].

\bibitem{chanowitz88}

M.S. Chanowitz,
Proceedingds: Workshop on Photon-Photon Collisions, Shores, Jerusalem Hills, 
Israel, April 24-28 (1988).


\bibitem{beveren02}
E. van Beveren et al., Mod. Phys. Lett. A \textbf{17}, 1673 (2002)
[hep-ph/0204139]; M.D. Scadron, G. Rupp, F. Kleefeld, E. van Beveren,
Phys. Rev. D \textbf{69}, 014010 (2004) [hep-ph/0309109].


\bibitem{jaffe77}
R.L. Jaffe, K. Johnson, Phys. Lett. B \textbf{60}, 201 (1976);
R.J. Jaffe, Phys. Rev. D \textbf{15}, 267; 281 (1977); 
\textbf{17}, 1444 (1978);
N.N. Achasov, V.V. Gubin, Phys. Rev. D \textbf{56}, 4084 (1997);
\textbf{63}, 094007 (2001); M. Alford, R.L. Jaffe, Nucl. Phys. B
\textbf{578}, 367 (2000); D. Black, A.H. Fariborz, J. Schechter,
AIP Conf. Proc. \textbf{549}, 241 (2002) [hep-ph/9911387]; F.E. Close,
N.A. T\"ornqvist, J. Phys. G: Nucl. Part. Phys. \textbf{28}, R249 (2002);
F.E. Close, Int. J. Mod. Phys. A \textbf{17}, 3239 (2002) [hep-ph/0110081];
R. Jaffe, F. Wilczek, Phys. Rev. Lett. \textbf{91}, 232003 (2003);
S.B. Gerasimov, Proceedings: 12th International Conference on Selected
Problems of Modern Physics (Blokhinsev03), Dubna,Russia, 8-11, June 200
[hep-ph/0311080]; 
F. Wilczek, in {\it From Fields to Strings}, edited by M. Shifman {et al.},
Vol \textbf{1}, pp 77-93 [hep-ph/0409168]; 
L. Maiani, F. Piccinini, A.D. Polosa, V. Riquer, Phys. Rev. Lett. 
\textbf{93}, 212002 (2004) [hep-ph/0407017];
Z.-G. Wang, W.-M. Yang, Eur. Phys. J. C \textbf{42},89 (2005)
[hep-ph/05501105], F. Giacosa,Phys. Rev. D \textbf{74}, 014028 (2006) 
[ hep-ph/0605191]. 

\bibitem{speth95}
G. Janssen, B.C. Pearce, K. Holinde, J. Speth, Phys. Rev. D \textbf{52},
2690 (1995); J. Speth, F. P. Sasson, S. Krewald, Nucl. Phys. A \textbf{721},
679 (2003).



\bibitem{achasov79}

N. N. Achasov, S. A. Devyanin, G. N. Shestakov, Phys. Lett. B \textbf{88}, 367
(1979).  

\bibitem{hanhart07}
C. Hanhart, B. Kubis, J. R. Pel\'aez, arXiv:0707.0262 [hep-ph]

\bibitem{wang04}

Z.-G. Wang, W.-M. Yang, S.-L. Wan, Eur. Phys. J. C \textbf{37}, 223 (2004)
[hep-ph/0410046] 

\bibitem{yao06}
W.-M. Yao et al. (Particle Data Group) J. Phys. C \textbf{33}, 1 (2006)
and 2007 partial update for edition 2008.


\bibitem{dalfaro73}

V. De Alfaro, S. Fubini, G. Furlan, C. Rossetti,  ``Currents in hadron
  physics'', North Holland (1973).





\bibitem{deswart63}
J. J. De Swart, Rev. Mod. Phys. \textbf{35}. 916 (1963).



\bibitem{erkol06}
G. Erkol, Doctoral Thesis, Groningen (2006);
G. Erkol, R.G.E. Timmermans, M. Oka, Th.A. Rijken, Phys, Rev, C \textbf{73},
044009 (2006).

\bibitem{dumbrajs83}
O. Dumbrajs {\it et al.}, Nucl. Phys. B \textbf{216}, 277 (1982).  


\bibitem{thomas00}

A.W. Thomas, W. Weise, ``The Structure of the Nucleon'', Wiley-VCH 
Berlin (2000).



\bibitem{armstrong72}
T.A. Armstrong {\it et al.}, Phys. Rev. D \textbf{5}, 1640 (1972);
Nucl. Phys. B \textbf{41}, 445 (1972).

\bibitem{arndt02}
R.A. Arndt {\it et al.}, Phys. Rev. C \textbf{66}, 055213 (2002).

\bibitem{hanstein98}
O. Hanstein, D. Drechsel, L. Tiator, Nucl. Phys. A \textbf{632}, 561 (1998).

\bibitem{drechsel99}
D. Drechsel, O. Hanstein, S.S. Kamalov, L. Tiator, Nucl. Phys. A \textbf{645},
145 (1999).

\bibitem{dutz05}

H. Dutz and the GDH collaboration, Phys. Rev. Lett \textbf{94}, 162001 (2005).


\end{thebibliography}
\end{document}